# TransONet: Automatic Segmentation of Vasculature in Computed Tomographic Angiograms Using Deep Learning

Alireza Bagheri Rajeoni
*Computer Science & Engineering*
*University of South Carolina*
Columbia, SC, USA
alirezab@email.sc.edu

Breanna Pederson
*Biomedical Engineering*
*University of South Carolina*
*School of Medicine*
Columbia, SC, USA
pedersob@email.sc.edu

Ali Firooz
*Computer Science & Engineering*
*University of South Carolina*
Columbia, SC, USA
ali.firooz@sc.edu

Hamed Abdollahi
*Computer Science & Engineering*
*University of South Carolina*
Columbia, SC, USA
ha25@mailbox.sc.edu

Andrew K. Smith
*Computer Science & Engineering*
*University of South Carolina*
Columbia, SC, USA
aks3@email.sc.edu

Daniel G. Clair, MD
*Surgery*
*Vanderbilt University*
Nashville, TN, USA
dan.clair@vumc.org

Susan M. Lessner
*Cell Biology & Anatomy*
*University of South Carolina*
*School of Medicine*
Columbia, SC, USA
susan.lessner@uscmed.sc.edu

Homayoun Valafar
*Computer Science & Engineering*
*University of South Carolina*
Columbia, SC, USA
homayoun@cse.sc.edu

*Abstract*—Pathological alterations in the human vascular system underlie many chronic diseases, such as atherosclerosis and aneurysms. However, manually analyzing diagnostic images of the vascular system, such as computed tomographic angiograms (CTAs) is a time-consuming and tedious process. To address this issue, we propose a deep learning model to segment the vascular system in CTA images of patients undergoing surgery for peripheral arterial disease (PAD). Our study focused on accurately segmenting the vascular system (1) from the descending thoracic aorta to the iliac bifurcation and (2) from the descending thoracic aorta to the knees in CTA images using deep learning techniques. Our approach achieved average Dice accuracies of 93.5% and 80.64% in test dataset for (1) and (2), respectively, highlighting its high accuracy and potential clinical utility. These findings demonstrate the use of deep learning techniques as a valuable tool for medical professionals to analyze the health of the vascular system efficiently and accurately. Please visit the GitHub page for this paper at https://github.com/pip-alireza/TransOnet.

*Keywords—Medical Image Segmentation, Computer Vision, Aorta, Artificial Intelligence, Peripheral Arterial Disease, Computed Tomography Angiogram*

## I. Introduction

The use of machine learning algorithms in the medical field has shown remarkable potential to aid healthcare professionals. For example, machine learning techniques have been shown to be advantageous in evaluating patient response to cardiac resynchronization therapy [1], sleep stage determination [2], prediction of patient response to medication administration [3], and hand and facial gesture recognition [4], [5]. The segmentation of the vascular system is another promising potential for advancement in medicine, as demonstrated by [6], where machine learning was used in a medical device to automatically locate peripheral vessels in ultrasound images . In this research paper, we propose a tailored deep learning algorithm designed to identify and extract the aorta and lower body arteries in CT scans as the first step in automating diagnostic and prognostic approaches by ML techniques.

Despite several attempts in segmenting the vascular system [7]–[9], a comprehensive analysis of the vascular system from the aorta to the knees, remains unexplored. The vascular system of the lower extremities carries crucial information about a person's health, as a complete blockage in these arteries can result in the need for amputation. Furthermore, analyzing the vascular calcification in the lower extremities may serve as a supplementary tool for assessing the risk of cardiovascular morbidity and mortality [3].

In this study, our objective is to trace the vascular system as it descends from the thoracic aorta, bifurcates into the iliac arteries, and further extends into the femoral arteries until it reaches the knee, as illustrated in Fig 1. The aorta, originating at the left ventricle, is the largest artery in the human body, running along the abdominal wall before branching into the common iliac arteries, which supply blood to the legs. Tracing the vascular system becomes increasingly challenging beyond the point of bifurcation as the size of the arteries diminishes, and they branch off multiple times, with the location of these smaller arteries varying from one patient to another. Additionally, blockages can occur in these arteries. It is worth noting that the area of interest also significantly decreases, often reducing to only a few pixels, making it challenging for machine learning models to accurately detect and segment the arteries, particularly if there is an occlusion.

Considering the complexities involved, our study aims to address these challenges and contribute to a better understanding of the vascular system in the lower extremities. By employing our application specific deep learning model, we strive to improve the accuracy of vascular tracing and segmentation beyond the point of bifurcation, which can have significant implications for diagnosis, treatment, and patient care.

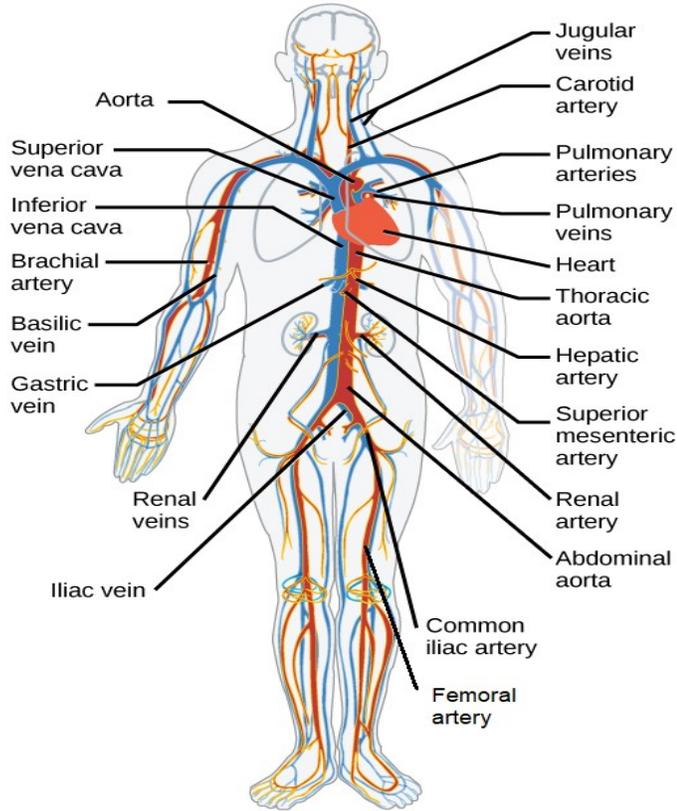

Fig. 1. Human vasculature map [10].

## II. Background and Methods

### A. Previous and Related Work

Several automated and semi-automated models have been proposed to segment CT scans to investigate organs such as lungs, abdominal aorta, liver, pancreas, and others [11]–[16]. Classically, machine learning techniques and threshold-based approaches were applied to hand-engineered features [11], [13]. Many of these approaches require expert human intervention. Human intervention including hand-engineering features is prone to introduce unrepresentative bias into the segmentation process. Further, modern deep learning systems have proven to be highly effective in learning complex nonlinear relationships, with much focus on image analysis [17]–[20]. For these reasons, it is desirable to develop an end-to-end deep learning approach for CT scan segmentation.

A few end-to-end deep learning approaches have been proposed to segment the aorta from CT scans. One such approach uses a dilated 3-dimensional convolutional neural network [21]. Others commonly use the UNet structure to segment the aorta from CT scans [22], [23]. Of the end-to-end deep learning approaches that segment CT scans to investigate bodily organs, none have specifically employed a block of Multi-Head Attention [24] between a pretrained ResNet encoder and decoder to segment aorta and lower body vessels. Multi-Head Attention is one of the best learning models in deep learning, lending itself strongly to computational parallelization.

One of the public challenges in this field is the Beyond The Cranial Vault (BTCV) [25], which aims to identify the best models for organ segmentation in medical image analysis. Some of the models that are used for segmentation include Swin UNETR [26], FCT [27], SwinUnet [28], nnFormer [29], and TransUnet [30], which utilize a combination of transformers and convolutional neural networks. However, there are only a few approaches focusing solely on vascular segmentation. Guidi et al [8] uses an expert system with convolutional neural networks to segment the vascular system from aorta to iliac arteries.

### B. Anatomical Background and Clinical Application

Peripheral arterial disease (PAD) is an atherosclerotic disorder of the arteries in the lower limbs. PAD is a common condition with over 200 million individuals estimated to have the condition globally. However, despite its prevalence, PAD is underdiagnosed, likely due to conflicting recommendations for screening, low awareness, and the diversity of symptoms as well as the number of asymptomatic cases [31]. Although it is a common cardiovascular disease, the treatment of PAD is often deficient when compared to other atherosclerotic diseases such as coronary artery disease. PAD is an important condition to screen for as it presages poor patient outcomes, cardiovascular morbidity, and mortality [31], [32]. Computed tomographic angiography (CTA) with intravenous contrast is a useful high-resolution imaging technique for visualizing pathological changes in the arterial tree. Its usefulness lies in the 3-dimensionality of the multidetector imaging, which allows for volumetric analysis, as the scans can be viewed in multiple 2-dimensional planes and as 3D reconstructions [33], [34]. In PAD treatment, CTAs are frequently used in pre-surgical planning to document the patient's vascular anatomy and to visualize the distribution of diseased, stenotic vessels. However, clinicians must take time during their office hours to do this annotation and it is not necessarily consistent. Machine learning (ML) algorithms and artificial intelligence (AI) systems are promising prospects for multiple fields of health care but are still in their infancy in clinical application [31]. For PAD, such techniques could be used to detect the disease, refine risk stratification, aid prognosis, and help with treatment choices [31], [32]. Some previous studies have applied AI/ML analyses to CTA scans of patients, demonstrating the potential to use CT imaging to visualize lower extremity inflow and runoff, integrate data in order to make predictions about the presence/absence of PAD, and predict risk of future cardiovascular events [32], [34]. There is emerging evidence that this type of model may predict risk better than conventional risk prediction scores based on linear modeling.

This study focuses primarily on the arteries that supply blood to the lower extremities, primarily the external iliac artery as it becomes the femoral artery and branches into the deep femoral artery in the leg. We begin tracking at the descending thoracic aorta as that is the largest artery in the body and supplies blood from the abdomen to the rest of the body. The study aims to evaluate the automated segmentation of the arterial system starting from the descending thoracic aorta throughout the relevant arteries in the leg to the knees. To evaluate the performance of the segmentation system, the data is annotated under the supervision of medical professionals, and the dataset is split into training, validation, and testing sets to assess the model's performance.

While there have been several approaches to the segmentation of the aorta by ML/AI techniques, there have been relatively fewer attempts at the segmentation of the arterial system past the iliac branching. In a recent study [35], convolutional neural networks (CNN) were employed to extract the vascular system and measure calcification, yielding promising results. In contrast, a prior work [36] utilized an object tracking technique in identifying the arterial system in the lower extremities. While successful in identifying the majority of the arterial system, it exhibited a shortcoming: if the object of interest was lost in one slice, or when intensity goes beyond the defined threshold, it would lose the object in all other subsequent slices of the CTA scan. This limitation was the primary cause of reduced performance in some instances.

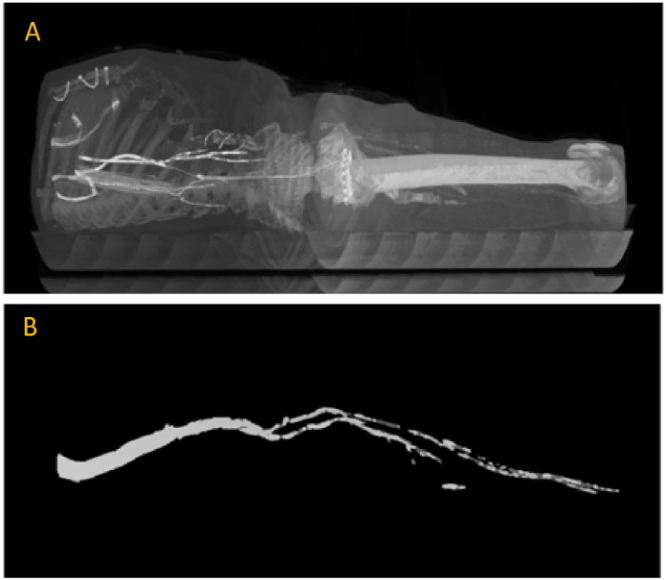

Fig. 2. A) is the 3D side view of human CT scan image. B) is the manually annotated vascular system.

### C. Data Description

The dataset consists of computed tomographic angiography (CTA) images, obtained with informed consent from 11 patients undergoing femoral endarterectomy for peripheral arterial disease at Prisma Health Midlands (IRB protocol 1852888). There are over 500 slices of images with height and width of 512 for each patient, extending from the descending thoracic aorta to the feet. Annotation is performed using ITK-SNAP [37] software under the supervision of a medical professional. ITK-SNAP provides a semi-automatic annotation tool that enables users to differentiate the area of interest from other regions by adjusting the intensity threshold. Users can create bubbles within the region of interest, and ITK-SNAP will automatically expand the selected region by following similar intensity patterns until the intensity drops at the edges. While ITK-SNAP offers a quick segmentation tool, it faces challenges when encountering complex geometries and vessel blockages, requiring more user intervention in these areas. Users must also carefully set the threshold for each patient and ensure there is no overlap between the area of interest and other regions having similar intensity.

An example of a dataset used for training is shown in Figs 2 and 3. Fig 2A shows a 3-dimensional reconstruction of the entire dataset, created using ImageJ software, with the vessels annotated using ITK-SNAP shown separately in Fig 2B. Fig 3A shows a typical transverse slice, with the annotated aorta in Fig 3B.

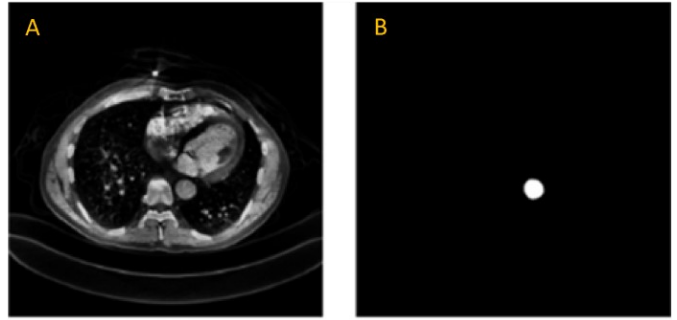

Fig. 3. A) Human CTA transverse view. B) Annotation of the aorta.

### D. Image Segmentation Method

Our model architecture follows an encoder-decoder structure similar to U-net [38], incorporating skip connections from the encoder to decoder. Additionally, a Transformer [39] is utilized in the bridge between these two main blocks as illustrated in Fig 4. As the network structure in Fig 4 resembles the letter "O" and to differentiate it from other networks, we name the architecture TransONet.

The encoder uses ResNet34 [19] with residual connections, from which four skip connections are extracted and fed to the decoder section. Given input $x$, the model accepts $x \in \mathbb{R}^{W \times H \times C}$ where $H$, $W$, and $C$ represents height, width, and the number of channels, respectively. To make the pretrained ResNet-34 model compatible with our dataset, the model's input is set to three channels. After feature extraction, the features undergo linear projection and reshaping. They are then passed through a bridge consisting layers of transformers. Subsequently, the features are reshaped back to ($H/3 \times W/3 \times 512$) and processed by a layer normalization and 2D convolutional layer. The purpose of this architecture is to utilize both high-resolution spatial information from CNN features as well as the global context encoded by Transformer. Transformers can serve as strong encoders for medical image segmentation tasks, with the combination of U-net to enhance finer details by recovering localized spatial information [30]. This custom-built structure utilizes pretrained ResNet-34 to improve training of deeper networks and solve vanishing gradients, especially when dealing with limited datasets. It allows leveraging the valuable initial feature representations learned from a larger dataset, which in turn contributes to improving the overall performance of the model.

The decoder, which has four blocks, receives the output of the encoder, and utilizes an expanding path to construct a segmentation map from the encoded features. Each decoding block duplicates the spatial resolution while halving the number of feature channels, performs upsampling, and concatenates the output with the symmetric block's output in the encoder. After that, the block applies a convolutional layer with the same number of filters. The output of the last decoder block undergoes a 1x1 convolution that calculates the final map. The structure of the TransONet model is shown in Fig 4. For developing TransONet we used Segmentation-Models and TensorFlow libraries.

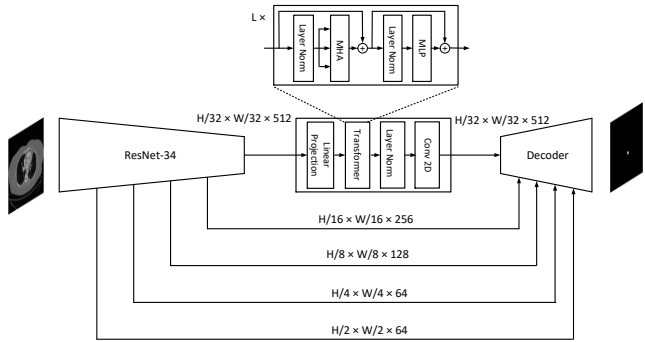

Fig. 4. TransONet structure. The decoder section utilizes ResNet-34. The output from the encoder undergoes linear projection and reshaping before passing through transformer. The transformed features are then reshaped and processed by layer normalization and 2D convolution. Subsequently, they are fed into the decoder to construct the segmentation mask. Skip connections from various stages of the encoder are sampled and incorporated into the decoder to contribute to the mask construction.

## III. RESULTS

To evaluate TransONet performance, we employed 4-fold cross-validation, as depicted in Fig 5. The training process involved using Binary Cross Entropy loss function and Adam [40] optimizer. The inputs to the model were images with H and W of 512 and three channels, and the output was a mask with the same size but one channel.

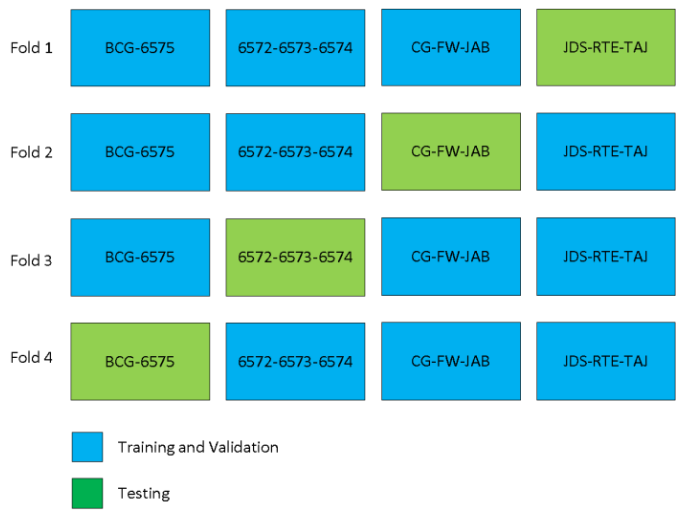

Fig. 5. Cross validation experiment. In every training fold, 3 patients are excluded, except in fold 4 where only two patients were excluded.

We trained the model using a batch size of 40, for 400 epochs. During training and validation, we utilized Binary Cross Entropy plus Jaccard (BCEJ) for loss and Intersection over Union (IOU) as the evaluation metric, while Dice score was used for testing. The results are shown in Fig 6.

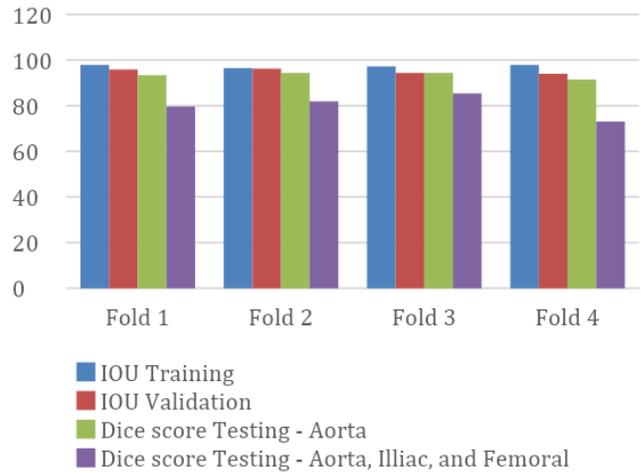

Fig. 6. Cross validation accuracy results.

Fig 7 illustrates the comparative analysis between our model and other architectures on our dataset. For training, we utilized BCEJ loss and IOU metric, conducting training for 200 epochs using a batch size of 10 and image dimensions of 256 in height and width. The graph in Fig 7 displays the IOU score during both training and validation stages. Notably, TransONet highlights superior performance on our in-house dataset.

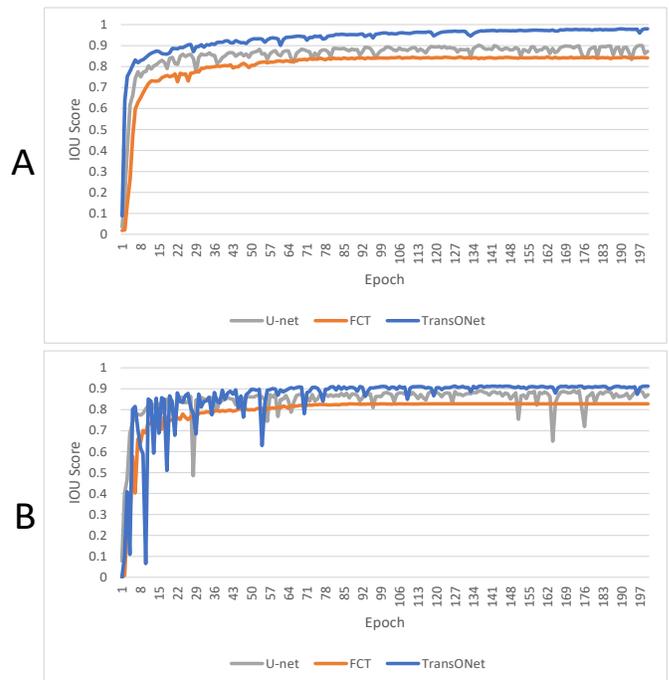

Fig. 7. Comparison of TransONet with FCT [27] and U-net [38] for aorta segmentation on our in-house dataset. Fig 7A illustrates the IOU score in training, and Fig 7B showcases the IOU score in validation across epochs. It is important to note that the validation set was entirely distinct from the training data.

Our model also achieved an average 80.6% Dice score in segmenting the vascular system from the descending thoracic aorta to the knee. Lower accuracy in this area is due to decreasing size of the arteries in the lower extremities. Additionally, contrast intensity drops due to stenoses in the lower body. In other words, the aortic cross-section in the upper body typically occupies more than 100 pixels, while the cross-sectional area of an artery in the leg may be less than 20 pixels.

In comparison to the object tracking model [36] using the same dataset, our approach demonstrate a significant improvement in vasculature segmentation. Fig 8 illustrates the results obtained by testing the model trained in fold 1 (Fig 6) on the RTE data. Compared to object tracking, TransONet demonstrated exceptional performance with a Dice score of 91.2% in segmenting the vascular system from the thoracic aorta up to the knee.

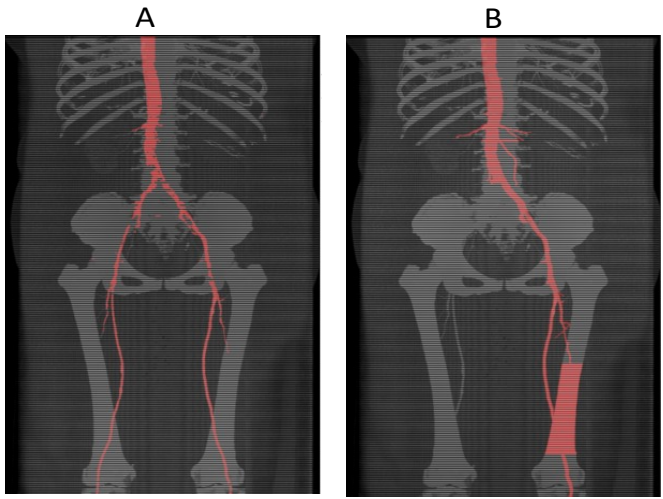

Fig. 8. Performance of TransONet 8A compared to object tracking 8B [36]. While object tracking methods may lose track of the vascular system or fail to avoid the skeleton, TransONet achieves successful tracking. However, TransONet may miss certain parts of the vascular system during the segmentation of each slice.

In contrast, object tracking clearly failed to accurately track the vascular system of the right leg. Object tracking follows the intensity and when in lower extremities blockages happen, and causes intensity to drop, object tracking loses the track of the vasculature. Also, in some cases the femoral arteries come in contact with bone structure and since they share similar intensity spectrum, object tracking identifies the bone structures as the vasculature which leads to the failure of the object tracking.

## IV. DISCUSSION AND FUTURE WORK

By accurately extracting and analyzing the vascular system, we can detect pathological conditions such as aneurysms and vascular calcification, among others. In the future, our focus will be on enhancing segmentation accuracy to precisely identify and measure calcification within the vascular system. This advancement will contribute to more accurate diagnosis, proactive treatment, and improved patient outcomes in the field of vascular health.

## ACKNOWLEDGMENTS

This work was funded by NIH grant number P20 RR-016461 to Dr. Valafar and HL145064-01 to Dr. Lessner. This work was also partially supported by the National Science Foundation EPSCoR Program under NSF Award # OIA-2242812.